\documentclass[journal]{IEEEtran}
\usepackage{cite}
\usepackage{amsmath,amssymb,amsfonts}
\usepackage{algorithmic}
\usepackage{graphicx,color}
\usepackage{textcomp}
\usepackage{hyperref}
\usepackage{subcaption}
\usepackage{comment}
\usepackage{booktabs}

\begin{document}

\title{Non-Blocking Fault Current Limiting Control of
Half-Bridge MMCs for MTDC Transmission}

\author{Pengxiang~Huang,~\IEEEmembership{Member,~IEEE},
and Shahil~Shah,~\IEEEmembership{Senior Member,~IEEE}%

\thanks{The authors are with National Laboratory of the Rockies (NLR), Golden, CO 80401 USA (e-mail: huangpx\_shiep@gwmail.gwu.edu; shahil.shah@nlr.gov).}%
\thanks{This work has been submitted to the IEEE for possible publication. Copyright may be transferred without notice, after which this version may no longer be accessible.}}

\maketitle

\begin{abstract}
A key protection requirement in multiterminal HVDC (MTDC) transmission systems is to selectively isolate faulted areas using DC circuit breakers (DCCBs) during DC faults while keeping the rest of the systems in operation. However, in MTDC systems using half-bridge modular multilevel converters (HB-MMCs), the lack of inherent fault current suppression capability in HB-MMCs can cause converter blocking before fault clearance by DCCBs. This work presents an active fault current limiting (AFCL) control strategy for HB-MMCs to mitigate DC fault currents, thus avoiding converter blocking and reducing the breaking capability required of DCCBs. In addition, a systematic approach is proposed to design the AFCL control strategy under converter and network stability constraints. The effectiveness of the proposed strategy is demonstrated through PSCAD on two typical MTDC systems during DC faults.
\end{abstract}


\section{INTRODUCTION}
\IEEEPARstart{M}{odular} multilevel converter (MMC)-based multiterminal high-voltage direct current (MTDC) transmission is rapidly expanding, driven by rising transmission demand as well as the power density, flexibility and reliability of MMC-MTDC transmission technology. MMCs employ either half-bridge (HB) or full-bridge (FB) submodules (SMs). HB SMs are favored for lower cost and losses but cannot generate negative voltage required to limit fault current during DC faults \cite{Pingyang2}. Thus, HB-MMCs block to protect IGBTs in SMs once the current flowing through IGBTs exceeds a threshold, and DC faults are cleared by opening AC-side breakers. Common in point-to-point HVDC, this strategy is not suited for MTDC systems, since the MMCs in the healthy parts of the networks shall continue their operation after a fault (i.e., selective protection) to minimize power flow interruptions and impact on the reliability of the bulk AC system \cite{Willem-DCCB-Sizing}. To enable selective protection of MTDC systems, DC circuit breakers (DCCBs) \cite{Hybrid-DCCB, PES-Magzine-DCCB}. However, the operation of DCCBs typically takes around 4 to 6 ms, while fault currents in HB-MMCs can reach the blocking threshold within a few hundred microseconds. This large mismatch implies many MMCs in an MTDC network can trip before fault isolation completes. Fault current limiting (FCL) methods are thus required for HB-MMCs of MTDC systems to limit fault currents and prevent converter blocking before fault isolation.

FCL methods can be categorized as passive and active. Passive methods employ extra hardware—such as current-limiting reactors (CLRs) \cite{Pang-Hui-Zhangbei-HVDC} and superconducting fault current limiters \cite{Bin-Li-SFCL}—to reduce the rate of rise of the fault current but increase cost and space demands. The integration of large CLRs can also reduce the response speed and stability margins of MTDC systems in both normal operation \cite{Cwikowski-Reactor-DC-Stability} and fault recovery process \cite{Willem-Post-Fault-Recovery, PH-Adaptive-FCL}. To maintain MTDC system stability, CLR size must stay below a threshold that can fail to prevent HB-MMC blocking during DC faults.

Active methods achieve FCL by reducing the number of inserted SMs per leg, that is, from $N$ to $\lambda N$ ($ 0\le \lambda < 1$). Methods in the existing literature differ in how and when $\lambda$ is set: either a fixed value of $\lambda$ is applied once a fault is detected, or $\lambda$ is adjusted dynamically by the active fault current limiting (AFCL) control. Refs. \cite{Full-bypass,Full-bypass-1} suggest bypassing all SMs (i.e., $\lambda$=0) immediately upon fault detection to force an exponential decay of fault current until DCCBs open, but this can cause severe arm overcurrent and trigger valve protection if DCCBs operation is slow. In contrast, \cite{Ke-Jia-Adaptive-FCL,Xiaoqian-Li-2} proposes bypassing a fixed amount of SMs ($0<\lambda<1$) at each stage of the fault current evolution (note that $\lambda$ varies between stages) to reduce the impact on MMC operation compared to \cite{Full-bypass, Full-bypass-1}. However, all these approaches require fault detection before executing SMs bypass, which typically takes 2–3 ms, and cannot ensure the MMC stays operational without blocking during this delay.
The studies presented in \cite{PH-AFCL-COMPEL, Rui-Li-Continuous-Operation-MTDC} introduced PID‐based fault-current control strategies that adaptively adjust $\lambda$ in response to fault–current severity, thereby adaptively regulating the number of inserted submodules (SMs) during faults. However, the PID design is inherently complex and insufficiently detailed in these papers, and its implementation may cause interactions between the MMC and the DC network.  

This paper proposes a novel AFCL strategy to prevent MMC blocking during DC faults in MTDC systems, enabling non-blocking DC fault ride-through (DC-FRT) while minimizing the size of passive CLRs. Existing AFCL methods typically require large CLRs (150–200 mH) \cite{Ke-Jia-Adaptive-FCL, Binye-Adaptive, Rui-Li-Continuous-Operation-MTDC}; the proposed method reduces them to 50 mH or less per MMC station. The AFCL strategy contains two parts:
\begin{itemize}
  \item Virtual-arm-impedance-based FCL (VAI-FCL) control — maps the determination of $\lambda$ into the circulating current feed-forward control and limits the rate of rise of the fault current. It remains active from MMC startup with negligible impact on normal operation, enabling immediate current-limiting when a DC fault occurs.
  \item Temporary-bypass-based FCL (TB-FCL) control — bypasses selected SMs once current exceeds a set threshold, ensuring exponential decay and capping the fault current peak, thus giving protection more time to isolate the fault and reducing DCCB breaking current.
\end{itemize}
To mitigate potential arm overcurrent caused by the AFCL, a virtual-resistance-based (VR-ACL) control is further introduced, emulating resistance at the MMC's AC terminal to suppress AC-side transient infeed and avoid MMC blocking due to arm overcurrent.

The remainder of this paper is organized as follows. Section \ref{Sec: II} analyzes DC fault current behavior under different numbers of inserted SMs. Section \ref{Sec: III} presents the proposed AFCL strategy. Section \ref{Sec: IV} gives AFCL design guidelines considering MMC stability and transient performance, and introduces the VR-ACL for limiting arm overcurrent. Section \ref{Sec: V} validates the strategy via PSCAD simulations of a radial six-terminal and a meshed four-terminal MTDC system under DC faults.

\section{DC Fault Current Behavior of Bipolar MMC}
\label{Sec: II}

\subsection{Generic Model of DC Fault Current of Bipolar MMC}
 \begin{figure}[!b]
    \centering
    \includegraphics[width=0.48\textwidth]{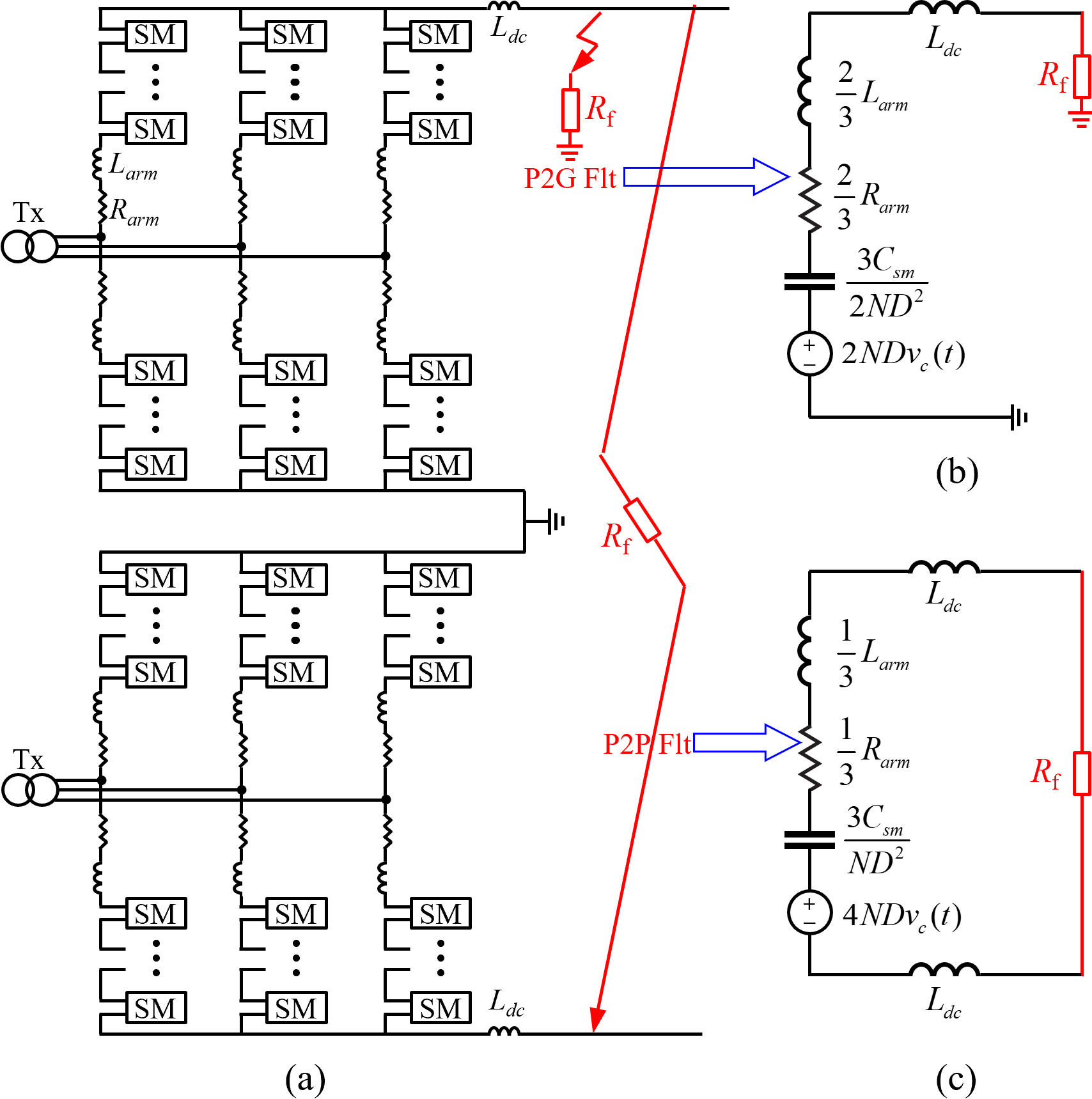}
    \caption{(a) Bipolar MMC with P2G and P2P fault applied at its DC terminals; equivalent circuit model of bipolar MMC during (b) P2G fault and (c) P2P fault.}
    \label{fig: Equivalent Circuit}
\end{figure}

As shown in Fig.~\ref{fig: Equivalent Circuit}(a), pole-to-ground (P2G) and pole-to-pole (P2P) faults are assumed at the DC terminal of the bipolar MMC, representing its most severe fault conditions. Following the modeling approach in~\cite{Xiaoqian-Li-1}, the state-space averaged fault circuits are depicted in Fig.~\ref{fig: Equivalent Circuit}(b) and (c), revealing that P2G and P2P faults share similar circuit structures. As P2G faults are considered the most probable in bipolar HVDC systems~\cite{Pingyang3}, this work focuses primarily on the P2G fault scenario. The DC fault current based on Fig.~\ref{fig: Equivalent Circuit}(b) can be expressed as:
\begin{flalign}
\label{eq: Fault Current}
L_{eq} C_{eq} \frac{\mathrm{d}^2 i_{dc}(t)}{\mathrm{d}t^2} 
+ R_{eq} C_{eq} \frac{\mathrm{d} i_{dc}(t)}{\mathrm{d}t} 
+ D^2 i_{dc}(t) = 0.
\end{flalign}

\noindent where $D$ denotes the equivalent duty ratio of the MMC, defined as the ration of ON-state SMs in an arm to the total number of SMs in that arm. $R_{eq}$, $L_{eq}$ and $C_{eq}$ are given by:
\begin{flalign}
\label{eq: Req, Leq and Ceq}
\begin{aligned}
R_{eq} = \frac{2}{3}R_{arm};\;
& \quad  L_{eq} = \frac{2}{3}L_{arm}+L_{dc};\;
& C_{eq} = \frac{3C_{arm}}{2N}.
\end{aligned}
\end{flalign}
The initial values for \eqref{eq: Fault Current} are given by \cite{Xiaoqian-Li-1}:
    \begin{flalign}
    \label{eq: Initial Value}
    &\hspace{-0em}
    \left\{
    \begin{aligned}
        i_{dc}(0^+) &= I_{dc}, \\
        \frac{\mathrm{d} i_{dc}(0^+)}{\mathrm{d}t} &= \frac{2D V_{dc} - R_{eq} I_{dc}}{L_{eq}}.
    \end{aligned}
    \right.
\end{flalign}

\noindent where $I_{dc}$ and $V_{dc}$ represent the DC current and the DC voltage right before the fault occurrence. The characteristic roots of \eqref{eq: Fault Current} are calculated by:
\begin{flalign}
\label{eq: Characteristic Roots}
\begin{aligned}
S_{1,2} = -\alpha  \pm \sqrt{\alpha^2 - \omega_0^2}.
\end{aligned}
\end{flalign}

\noindent where $\alpha =R_{eq}/(2L_{eq})$ and $\omega_0 = D/\sqrt{({L_{eq}C_{eq})}}$. Observed from \eqref{eq: Characteristic Roots}, the solution of $i_{dc}(t)$ in \eqref{eq: Fault Current} is determined by the difference between $D$ and $R_{eq}/(2\sqrt{L_{eq}/C_{eq})}$; thus, $i_{dc}(t)$ is solved based on different selections of $D$.

\subsection{Responses of \texorpdfstring{$i_{dc}(t)$}{idc} Under Different Values of \textit{D}}


\subsubsection{\texorpdfstring{$ R_{eq}/(2\sqrt{L_{eq}/C_{eq}}) < D$}{TAO<D}}

Owing to the typically small $R_{eq}$ in MMC and DC systems, $D$ commonly falls within this range under fault conditions and can be further categorized as:
\begin{itemize}
    \item $D = 0.5$: When only CLR is used for current-limiting, $D$ remains unchanged from its steady-state value within the first 2 to 4 ms after fault occurs. 
    \item $ R_{eq}/(2\sqrt{L_{eq}/C_{eq}}) < D < 0.5$: When AFCL control is applied, $D$ can be reduced to a value smaller than $0.5$ but larger than $R_{eq}/(2\sqrt{L_{eq}/C_{eq}})$. 
\end{itemize}

\noindent $i_{dc}(t)$ under these two scenarios is solved as \cite{Xiaoqian-Li-2}:
\begin{flalign}
\label{eq: Idc Solution 1}
\begin{aligned}
i_{dc} (t) = A e^{-\alpha t} \sin\left(\omega_r t + \beta \right).
\end{aligned}
\end{flalign}
in which: 
\[
\left\{
\begin{aligned}
A &= \frac{I_{dc}}{\sin \beta}, \quad
\beta = \arctan\left(\frac{2\omega_r L_{eq} I_{dc}}{4D V_{dc} - R_{eq} I_{dc}}\right), \\
\omega_r &= \sqrt{\left| \omega_0^2 - \alpha^2 \right|}, \quad 
\gamma = \arctan\left(\frac{\omega_r}{\alpha}\right)
\end{aligned}
\right.
\]

Because the rate of rise of the fault current is primarily influenced by $\omega_r$ in \eqref{eq: Idc Solution 1}, a smaller $D$ reduces this rate and effectively limits the rise of the fault current.
\subsubsection{\texorpdfstring{$0 < D < R_{eq}/(2\sqrt{L_{eq}/C_{eq}} )$}{0<D<Tao}}

The value of $D$ falls within this range under one of the following conditions:
\begin{itemize}
    \item The MMC is equipped with a resistance-type fault current limiter \cite{Lei-Chen-RSFCL}, which increases $R_{eq}$ during a DC fault.
    \item Some AFCL controls can lead to a relatively small value of $D$ (excluding the case when $D = 0$).
\end{itemize}

\noindent $i_{\text{dc}}(t)$ in such a condition of $D$ is expressed as:
\begin{equation}
    i_{dc}(t) = A_1 e^{S_1 t} + A_2 e^{S_2 t}
    \label{eq: Idc Solution 2}
\end{equation}
\begin{flalign}
\label{eq: A1 and A2 for Idc Solution 2}
\begin{aligned}
A_{1,2} = (2\omega_r L_\text{eq} I_\text{dc}  \pm  4DV_\text{dc} \mp R_\text{eq} I_\text{dc})/(4\omega_r L_\text{eq})
\end{aligned}
\end{flalign}
\subsubsection{\texorpdfstring{$D=0$}{D=0}}

This condition arises only when all SMs are bypassed, which results in the capacitor impedance and the voltage source in Fig. \ref{fig: Equivalent Circuit}(b) being reduced to zero. As a result, $i_{{dc}}(t)$ is:
\begin{flalign}
\label{eq: Idc Solution 3}
\begin{aligned}
i_{dc}(t) = I_{dc} e^{-2\alpha t}
\end{aligned}
\end{flalign}
\noindent As can be seen, the response of $i_{dc}(t)$ exhibits a first-order exponential decay behavior.

Equation \eqref{eq: Idc Solution 1} shows that $D$ governs the rising rate of the fault current right after the fault occurs but does not constrain its peak value. In contrast, \eqref{eq: Idc Solution 2} and \eqref{eq: Idc Solution 3} describe the fault current in attenuation behavior, which can be used to limit the fault current maximum. These expressions, however, are derived based on the assumption that the DC fault occurs at the MMC terminal. 
When long transmission lines or DC networks with complicated topologies exist between the fault location and the MMC, analytical derivation of $i_{dc}(t)$ becomes difficult or impractical, and the effects of $D$ and fault locations on the behavior of $i_{dc}(t)$ must be assessed via offline simulations. Nevertheless, any protection design, including AFCL, typically considers the worst-case condition—a terminal fault—so\eqref{eq: Idc Solution 1}–\eqref{eq: Idc Solution 3} remain valuable for elucidating how adjustments of $D$ can shape $i_{dc}(t)$.

\section{Active Fault Current Limiting Strategy}
\label{Sec: III}
\subsection{Virtual Arm Impedance-Based FCL Control (VAI-FCL)}

The relationship between duty cycle $D$, the arm voltages, and the DC-bus voltage satisfies:
\begin{flalign}
\label{eq: Expression of D}
\begin{aligned}
D =\frac{v_u^j(t) + v_l^j(t)}{2 v_{dc}(t)}
\end{aligned}
\end{flalign}

\noindent where $v_u^j(t)$ and $v_l^j(t)$ represent the upper and lower arm voltages of phase $j$ ($j = a, b, c$).  According to \cite{Qingrui-Tu-CCSC}, $v_u^j(t)$ and $v_l^j(t)$ can be expressed as follows:
\begin{flalign}
\label{eq: Expression of V_xu}
\begin{aligned}
v_u^j(t) = V_{dc}/2 - v_{s}^{j*}(t)-v_z^{j*}(t)
\end{aligned}
\end{flalign}
\begin{flalign}
\label{eq: Expression of V_xl}
\begin{aligned}
v_l^j(t) = V_{dc}/2 + v_{s}^{j*}(t)-v_z^{j*}(t)
\end{aligned}
\end{flalign}

\noindent where $v_{s}^{j*}(t)$ represents the phase voltage reference generated by the upper-level controller (e.g., phase current controller), and $v_z^{j*}(t)$ is the common-mode voltage reference regulated by the circulating current controller. As observed from \eqref{eq: Expression of D}, \eqref{eq: Expression of V_xu} and \eqref{eq: Expression of V_xl}, the value of $D$ can be reduced by enlarging $v_z^{j*}(t)$. To achieve this without modifying the existing circulating current control design, an additional circulating current feed-forward loop can be introduced.
 \begin{figure}[!t]
    \centering
    \includegraphics[width=0.5\textwidth]{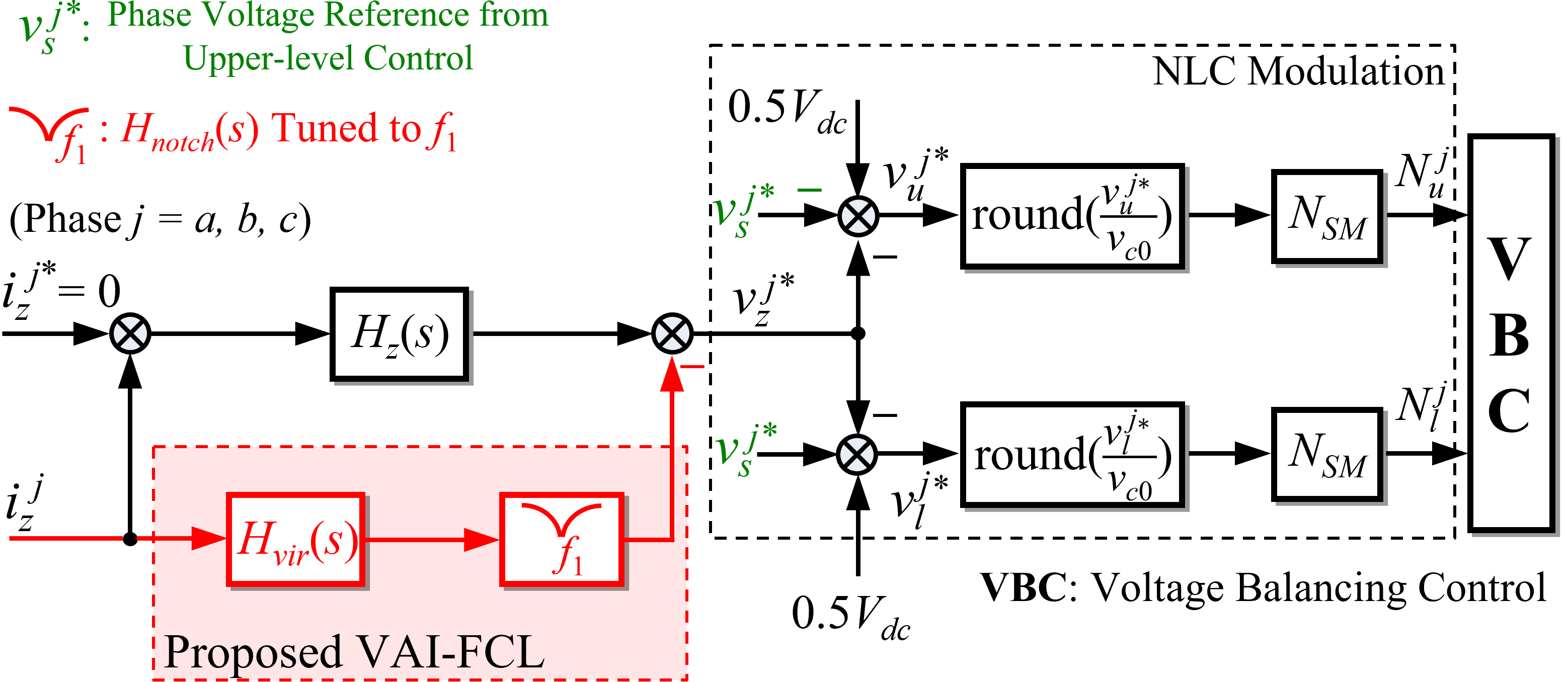}
    \caption{Control structure of VAI-FCL control.}
    \label{fig: VAI-FCL}
\end{figure}
In Fig.~\ref{fig: VAI-FCL}, the area marked by the red box represents the block diagram of the VAI-FCL. $i_z^j$ represents the circulating current in phase $j$ ($j = a, b, c$); and $H_z(s)$ represents the $2^{nd}$-order circulating current-suppressing control (CCSC) \cite{Qingrui-Tu-CCSC}. As shown, VAI-FCL uses the circulating current as its input, and its output is subtracted from the modulation reference generated by the CCSC, $v_z^{j*}$. This is equivalent to inserting a virtual series impedance into the arm inductor, where the impedance characteristics of this virtual element are defined by the control functions of the VAI-FCL. Since the primary objective of this virtual impedance is to limit the rate of rise of the fault current, the control function of the VAI-FCL is selected as a derivative term, $sL_v$. In addition, to extract the DC component while mitigating potential interactions between the CCSC and the feed-forward loop—and to suppress the undesired amplification of high-frequency components by the derivative term—a first-order low-pass filter with a low cutoff frequency (denoted as $f_\text{c}$) shall be incorporated into the VAI-FCL loop. As a result, the transfer function of the VAI-FCL controller, $H_{vir}(s)$, is given by: 
\begin{flalign}
\label{eq: Transfer Function of H_vir}
\begin{aligned}
H_{vir}(s)  = sL_v\frac{2\pi f_c}{s+2\pi f_c}
\end{aligned}
\end{flalign}

It should be noted that the CCSC does not directly regulate the arm energy \cite{Shenghui-Cui-Energy}, which may lead to a slight energy imbalance between the upper and lower arms of the MMC. Such an imbalance generally induces a negligible circulating current at the fundamental frequency. However, depending on the specific parameter design, the gain of $H_{vir}(s)$ may not be sufficiently low at the fundamental frequency. For instance, when $L_v = 0.2$ and $f_c = 5$ Hz, the magnitude of $H_{vir}(s)$ at 60 Hz is 6.26. Consequently, a small 60-Hz component can be amplified, causing distortion in the arm current. This issue can be effectively mitigated by incorporating additional notch filters tuned to the fundamental frequency, as illustrated in Fig. \ref{fig: VAI-FCL}. The transfer function of the notch filter is given as:
\begin{flalign}
\label{eq: Transfer Function of H_notch}
\begin{aligned}
H_{notch}(s) = \frac{s^2 + (2 \pi f_1)^2}{s^2 + 2 \pi f_c s + (2 \pi f_1)^2}
\end{aligned}
\end{flalign}

\noindent where $f_c = 6$ Hz is selected to accommodate a potential variation of $\pm 3$ Hz in the fundamental frequency.

Note that under steady-state operating conditions, the DC current in each arm remains approximately $I_{dc}/3$, where $I_{dc}$ denotes the rated DC current. As a result, the output of VAI-FCL is virtually zero, indicating that it does not affect the steady-state operation of the MMC. Although the steady-state DC current may vary slightly due to power fluctuations or system adjustments, the corresponding rate of change is typically small; thus, the VAI-FCL's output remains negligible and has no practical influence on MMC operation. In contrast, once a fault causes a rapid and significant increase in DC current, the virtual inductor, $L_v$, is automatically introduced into each arm to limit the rate of rise of the fault current. This mechanism provides an adaptive current-limiting capability that operates independently of fault detection or discrimination by the protection system. 

%
\subsection{Temporary SM Bypass-Based FCL Control (TB-FCL)}

As indicated by \eqref{eq: Idc Solution 2} and \eqref{eq: Idc Solution 3}, the DC current exhibits an exponential decay when $D$ satisfies $0 \le  D < R_{eq}/(2\sqrt{L_{eq}/C_{eq}})$. Leveraging this property, this subsection introduces the TB-FCL that directly limits the fault current peak to further enhance the FCL capability of the HB-MMC.

\begin{figure}[!t]
    \centering
    \includegraphics[width=0.5\textwidth]{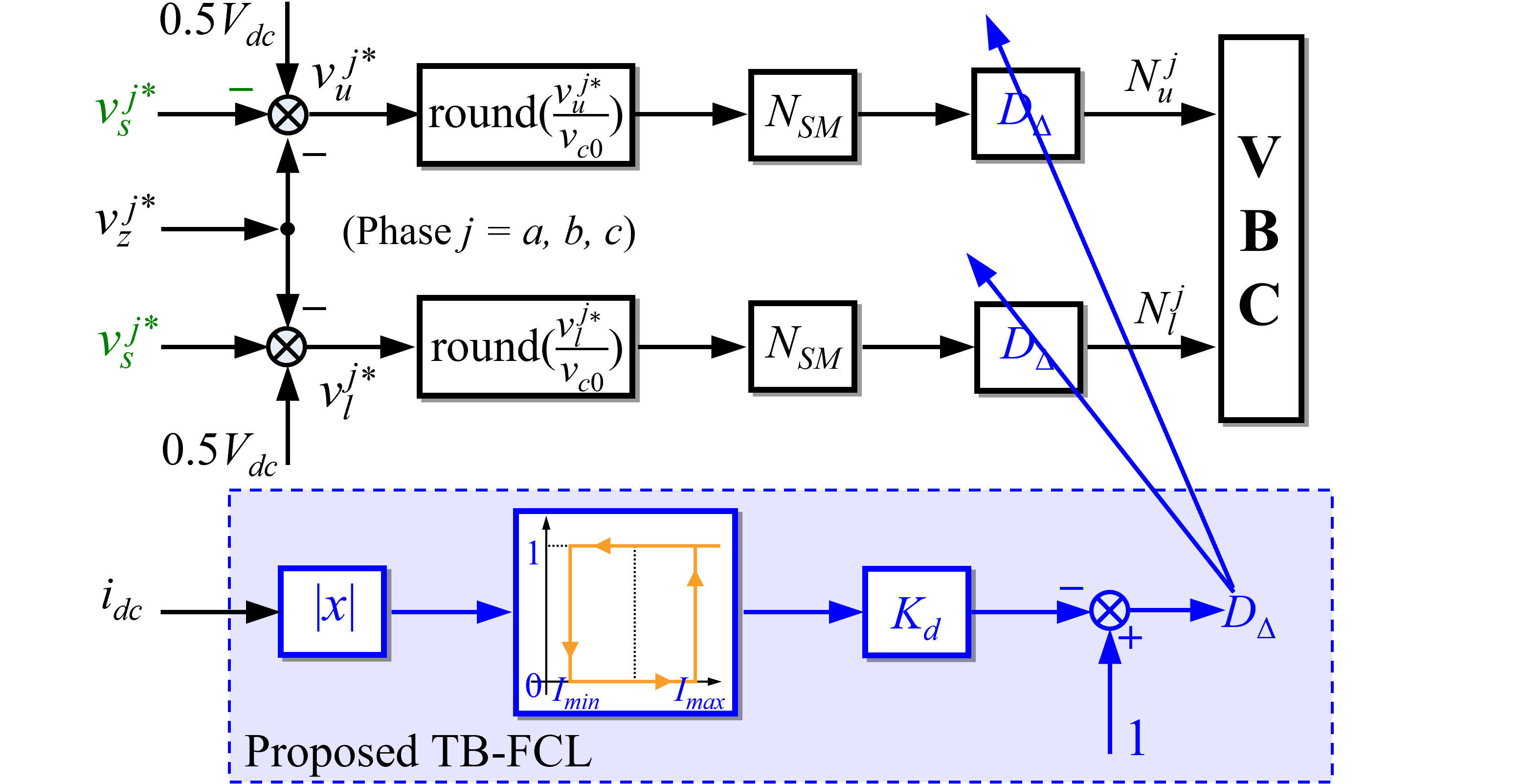}
    \caption{Control structure of TB-FCL control.}
    \label{fig: TB-FCL}
\end{figure}

The blue-dashed box in Fig. \ref{fig: TB-FCL} illustrates the configuration of the TB-FCL, where $i_{dc}$ represents the DC-bus current of the MMC. This current is processed through a nonlinear gain block that generates a binary output signal, switching between 0 and 1 within a hysteresis band defined by the thresholds $I_{max}$ and $I_{min}$. The output of the nonlinear gain block is then multiplied by $K_d$ and subtracted by $1$, yielding a modulation factor, $D_\Delta$. This factor adjusts the number of SMs inserted per arm, modifying it from $N_{SM}$ to $D_\Delta N_{SM}$ during each switching cycle. The parameter $K_d$ can be selected within the range of $1- R_{eq}/(2\sqrt{L_{eq}/C_{eq}})$ to $1$.
During the startup and steady-state operation of the MMC, even with a power fluctuation, the amplitude of $i_{dc}$ should always remain below $I_{max}$, ensuring that the output of the nonlinear gain block remains $0$. Consequently, $D_\Delta $ is maintained at $1$, meaning that the TB-FCL does not affect the normal switching of the MMC. In the event of a DC fault once $i_{dc}$ exceeds $I_{max}$, however, the nonlinear gain block outputs $1$, causing the modulation factor to switch to $1-K_d$. For instance, when $K_d = 1$, $D_\Delta = 0$ and all SMs are bypassed, the DC fault current responses is changed from that described in  \eqref{eq: Idc Solution 1} to that in \eqref{eq: Idc Solution 3}, indicating an immediate decay of the fault current. Once $i_{dc}$ decreases below $I_{min}$, the output of the nonlinear gain block switches back from 1 to 0, allowing $N_{SM}$ SMs to be inserted per switching cycle. If the DC fault remains not cleared, $i_{dc}$ will increase again, triggering the TB-FCL whenever $|i_{dc}|$ exceeds $I_{max}$.

\subsection{Sequence of Events WO/ and W/ the AFCL Strategy}
\begin{figure}[!t]
    \centering
    \includegraphics[width=0.5\textwidth]{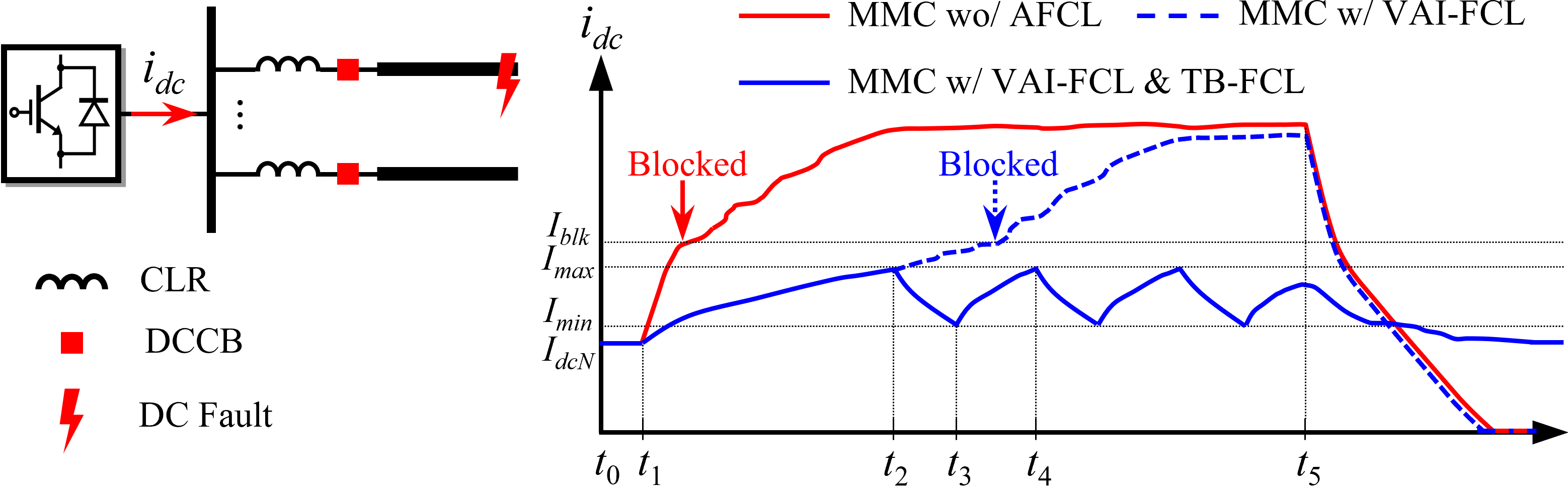}
    \caption{Progress of the MMC protection with the proposed AFCL strategy (VAI-FCL and TB-FCL).}
    \label{fig: Protection Sequence}
\end{figure}

The protection sequence when using the proposed AFCL strategy is shown by the solid-blue line in Fig. \ref{fig: Protection Sequence}.
\begin{itemize}
    \item $t_0$: The fault occurs in the DC network and creates a voltage and current wave traveling toward the MMC station. 
    \item $t_1$: The traveling wave propagates to the DC bus of the MMC, 
    and the DC fault detection is activated to discriminate the fault.
    \item $t_1 \sim t_2$: The DC current, $i_{dc}$, experiences a significant increase due to the discharge of the MMC SMs, causing the VAI-FCL to immediately insert virtual arm inductors into all six arms, thereby limiting the rate of rise of $i_{dc}$. 
    \item $t_2$: $i_{dc}$ reaches $I_{max}$, and the TB-FCL is activated to limit the peak value of $i_{dc}$.
    \item $t_2 \sim t_3$: Due to the bypass of a large amount of the SMs, $i_{dc}$ decreases. 
    \item $t_3$: $i_{dc}$ decreases to $I_{min}$, the TB-FCL is deactivated, and the all SMs resume normal insertion and bypass modes.
    \item $t_3 \sim t_4$: Because the DC fault is not cleared yet, the SM continues to discharge, resulting in an increase in the DC fault current. This causes the VAI-FCL to reinsert virtual reactors into the six arms.
    \item $t_4 \sim t_5$: The VAI-FCL and the TB-FCL operate alternately in a repetitive manner until the fault is cleared.
    \item $t_5$: The faulted area is isolated from the system by the DCCBs, and $i_{dc}$ starts to decrease and recover to its steady-state value.
\end{itemize}

It should be noted that if the fault is cleared before time instant $t_2$,  the TB-FCL control may not be triggered. To further explain the importance of TB-FCL control, a comparison of the fault current responses in Fig.~\ref{fig: Protection Sequence}—including the MMC without AFCL strategy (solid red line) and the MMC equipped solely with the VAI-FCL control (dashed blue line)—reveals the following observations: (1) without AFCL strategy, the MMC is blocked immediately after the fault occurs, and the fault current increases to an AC steady-state infeed; (2) if the fault is not cleared promptly, the MMC may still be blocked, even when the VAI-FCL is employed.

\section{Design Consideration of the AFCL Strategy}
\label{Sec: IV}
\subsection{Design of VAI-FCL Control} 

\subsubsection{Controller Stability-Oriented Design} 
\begin{figure}[!t]
    \centering
    \includegraphics[width=0.5\textwidth]{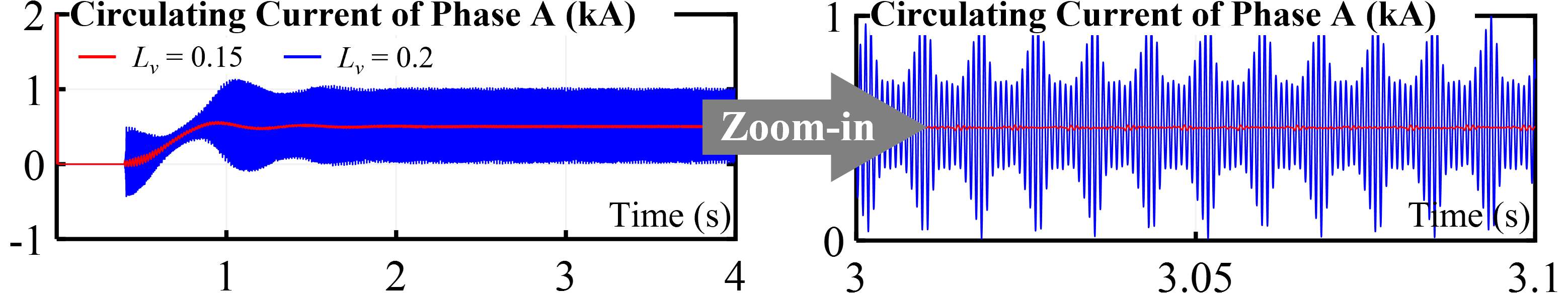}
    \caption{Examples demonstrating the stability of the circulating current control when selecting: $L_v = 0.15$ (red) and $L_v = 0.2$ (blue).}
    \label{fig: Simulated Responses of CC}
\end{figure}

As shown in Fig.~\ref{fig: VAI-FCL}, the VAI-FCL control is equivalent to adding a derivative term to the existing circulating current control, which can impact the stability of the circulating current control. In addition, a control delay of MMC can reduce the phase margin of the VAI-FCL control, further jeopardizing the stability of the circulating current loop. As a result, controller stability must be carefully considered when selecting $L_v$. To explain the control stability issue that is the concern of this subsection, simulated responses of an MMC connected to an ideal voltage source are presented in Fig.~\ref{fig: Simulated Responses of CC}. The MMC uses the designs of MMC C-2 presented in Table~\ref{Tab: MMC Parameter in Case 2}, and a $200~\mu$s control delay is included. As shown in Fig.~\ref{fig: Simulated Responses of CC}, when $L_v$ is changed from  $0.15$ to $0.2$, the circulating current response contains a stable oscillation after the MMC starts at $t  = 0.4$ s. 

\begin{figure}[!t]
    \centering
    \includegraphics[width=0.5\textwidth]{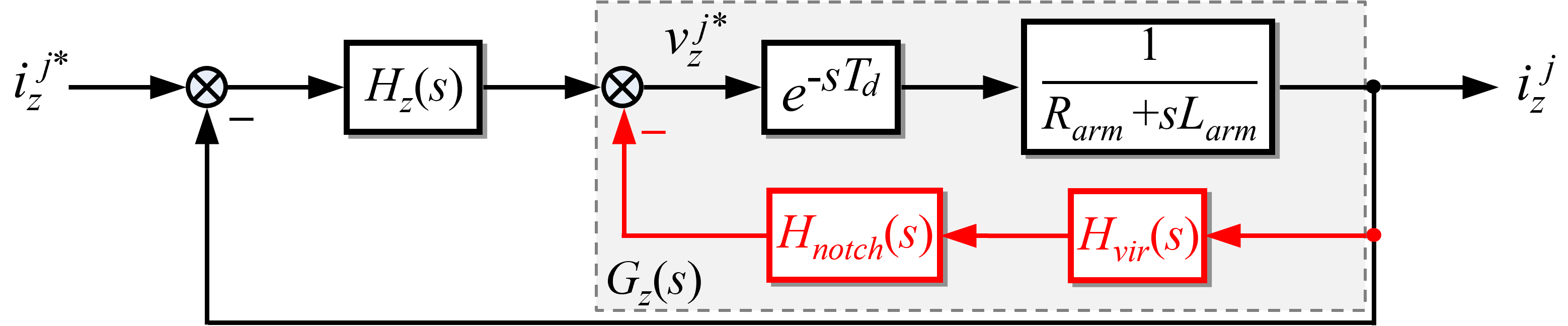}
    \caption{Block diagram representation of the circulating current control with VAI-FCL.}
    \label{fig: Block Diagram of CCC w/ VAI-FCL}
\end{figure}

Based on Fig.~\ref{fig: VAI-FCL}, \eqref{eq: Transfer Function of H_vir}, and \eqref{eq: Transfer Function of H_notch}, Fig.~\ref{fig: Block Diagram of CCC w/ VAI-FCL} depicts the block diagram representation of the circulating current control with VAI-FCL control, and the closed-loop transfer function of the circulating current loop is given by:
\begin{flalign}
\label{eq: Loop Gain of CCC}
\begin{aligned}
P_{z}(s)  = \frac{H_z(s)G_z(s)}{1+H_z(s)G_z(s)}
\end{aligned}
\end{flalign}

\noindent where $G_z(s)$ is: 
\begin{flalign}
\label{eq: Transfer Function of Gz(s)}
\begin{aligned}
G_z(s) = \frac{e^{-sT_d}/(sL_{arm}) }{1+H_{notch}(s)H_{vir}(s)e^{-sT_d}/(sL_{arm}) }
\end{aligned}
\end{flalign}

\noindent The application of the Hurwitz criterion to \eqref{eq: Loop Gain of CCC} is necessary to determine whether the circulating current loop remains stable for a given selection of $L_v$. In general, a smaller $L_v$ has a reduced impact on the circulating current controller. Consequently, the controller stability-oriented design provides the maximum value of $L_v$ (denoted as $L_{v,max}$) while ensuring the stable operation of the circulating current loop.
\subsubsection{System Stability-Oriented Design}
Note that the selection of $L_v = L_{v,max}$ shall not compromise system stability also. According to the impedance-based stability theory for HVDC grid \cite{DC-harmonic-resonance}, the ratio $Z_g(s)/Z_{dc}(s)$ must satisfy the Nyquist criterion, where $Z_g(s)$ and $Z_{dc}(s)$ denote the DC impedance of the HVDC grid and the MMC, respectively. 
 \begin{figure}[!t]
    \centering
    \includegraphics[width=0.5\textwidth]{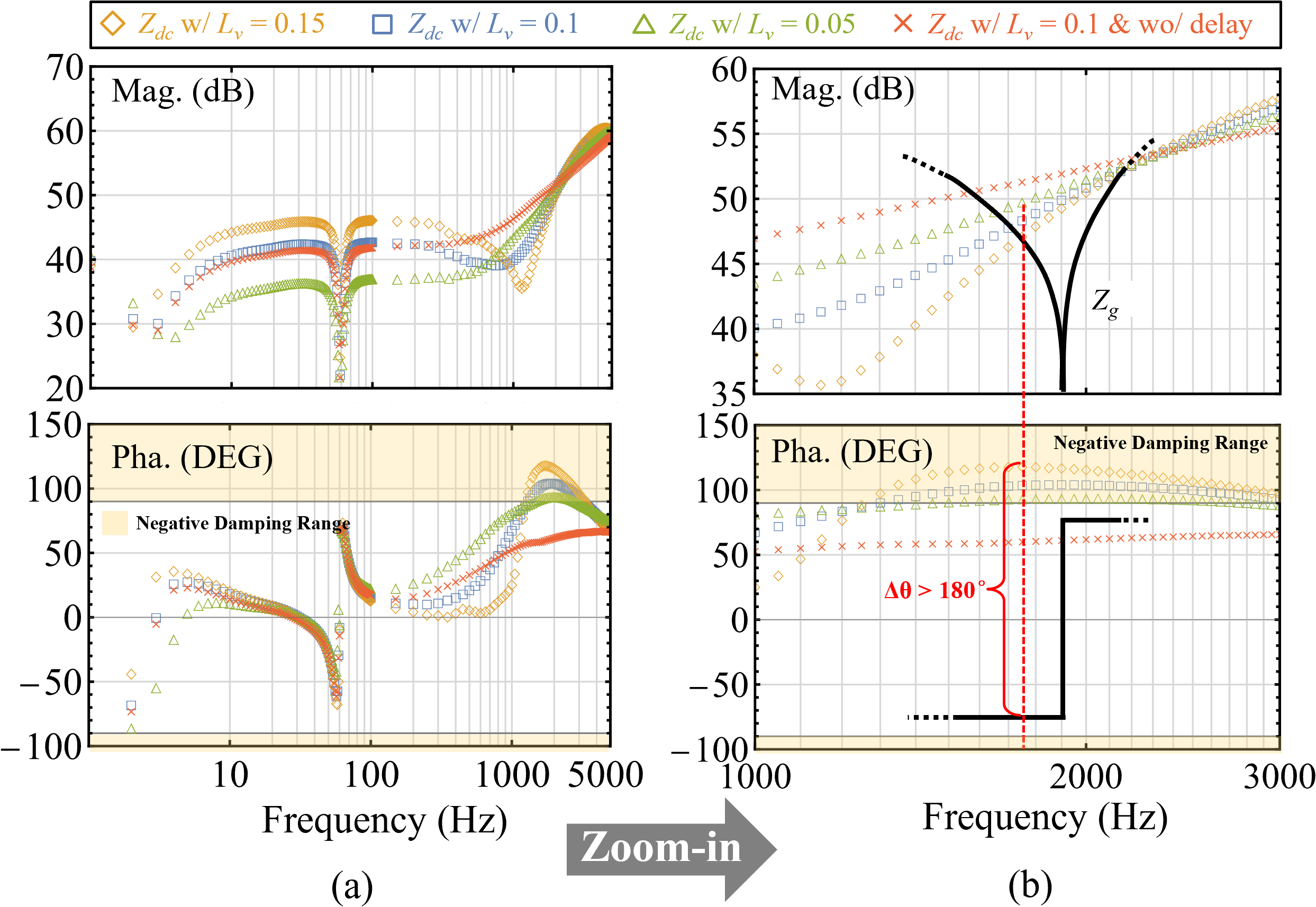}
    \caption{Effetcs of the VAI-FCL on the DC impedance of the PQ-controlled MMC with a $200~\mu$s delay.}
    \label{fig: VAI-FCL with Different Lv}
\end{figure}
To illustrate the effect of $L_v$ on the DC impedance of MMC, Fig.~\ref{fig: VAI-FCL with Different Lv} plots the frequency scan result of the DC impedance responses of the MMC with $200~\mu$s of control delay (denoted as $T_d = 200~\mu$s) when $L_v$ is increased from 0.05 to 0.15 in a 0.05 step. The without-delay case is also included for comparison. It is observed that the phase of $Z_{dc}(s)$ remains within the range from -90° to 90° for frequencies below $\sim 1.5$ kHz,regardless of the selection of $L_v$. This indicates no subsynchronous and supersynchronous resonances, nor any high-frequency resonance in the HVDC grid during steady-state operation; however, the control delay introduces negative damping on the DC impedance responses of the MMC at frequencies between $\sim 1.5$ kHz and $\sim 3.75$ kHz. This could result in instability between the MMC and the HVDC grid in this frequency range, depending on the amount of positive damping of the HVDC grid. On the other hand, reducing the value of $L_v$ decreases the amount of negative damping of MMC DC impedance, which could help mitigate the system resonance within this frequency range. To further illustrate the instability issue between the MMC and the HVDC grid due to the improper design of the VAI-FCL, Fig.~\ref{fig: VAI-FCL with Different Lv}(b) depicts an example HVDC grid impedance around 2 kHz (black solid line) and compares it against the DC impedance responses of the MMC in Fig.~\ref{fig: VAI-FCL with Different Lv}(a). As shown, $Z_g(s)$ intersects with $Z_{dc}(s)$ at approximately 1750 Hz when $L_v = 0.15$. And the phase difference between the two exceeds 180°, indicating an unstable resonance condition. Reducing $L_v$ to $0.1$ or $0.05$ effectively mitigates this instability by preventing unstable resonance conditions between $Z_g(s)$ and $Z_{dc}(s)$. For more detailed analysis on how the negative damping of the DC impedance of MMC interacts with the HVDC grid impedance and results in instability, please refer to \cite{DC-harmonic-resonance}.

In a practical HVDC system, when $Z_g(s)/Z_{dc}(s)$ does not meet the Nyquist stability criterion, reducing the value of $L_v$ becomes necessary to ensure system-level stability; however, this might impact the current-limiting capability provided by the VAI-FCL control. 
\subsubsection{Iterative Design Process for Selecting \texorpdfstring{$L_v$}{Lv}}

\begin{figure}[!t]
    \centering
    \includegraphics[width=0.51\textwidth]{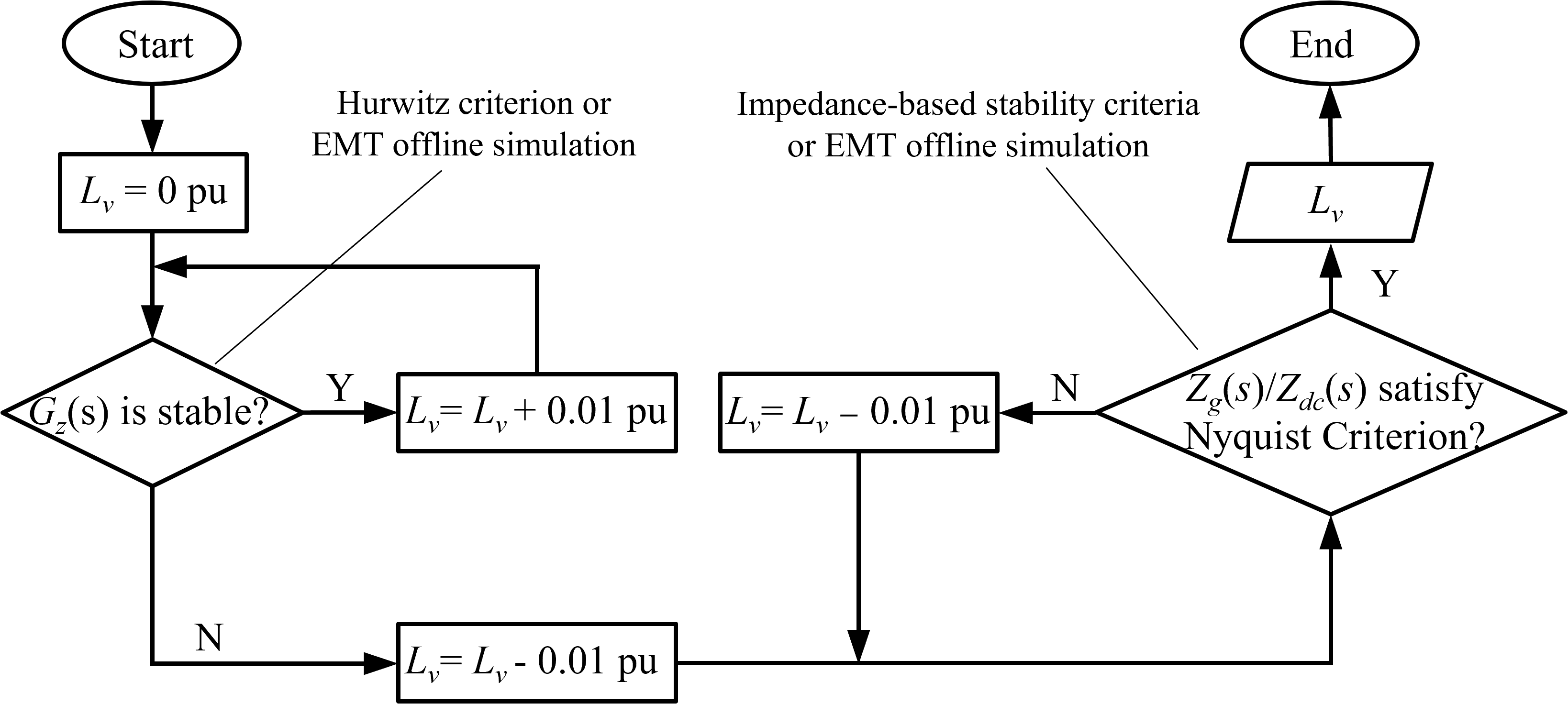}
    \caption{Iterative design process for $L_v$.}
    \label{fig: Iterative Process}
\end{figure}

To achieve a satisfactory design of $L_v$ considering both controller and system stability, an iterative design process is proposed here. A flowchart summarizing the process is depicted in Fig.~\ref{fig: Iterative Process}.
Note that when evaluating the stability of $G_z(s)$ and system, in addition to employing the Hurwitz and impedance-based stability criteria, offline electromagnetic transient simulations can be used to identify inappropriate selections of $L_v$ that could lead to instability. This is more efficient and straightforward, especially when the vendor-supplied white-box or black-box model is available.

\subsubsection{Suppressing Arm Overcurrent Due to SM Bypass}
The VAI-FCL control is essentially an SM bypass scheme in which the number of bypassed SMs is adaptively determined by the feed-forward controller, $H_{vir}(s)$, based on the severity of the discharge of the MMC SM capacitors. Because the VAI-FCL control operates immediately upon the fault occurrence, the modulator gain of the MMC remains unchanged from its steady-state value. This inevitably results in a decrease of the MMC’s AC output voltage. In a weak AC system—such as one with a long overhead transmission line—or in an MMC-based renewable energy integration system (AC current controlled by a renewable energy plant), this reduction in AC output voltage typically does not result in a significant increase in AC current. On the contrary, when the MMC is connected to a strong AC system, a reduction in the AC output voltage of the MMC can lead to a significant increase in the AC current, potentially resulting in SM blocking due to the valve overcurrent protection \cite{Valve-current-protection}. 
 \begin{figure}[!b]
    \centering
    \includegraphics[width=0.5\textwidth]{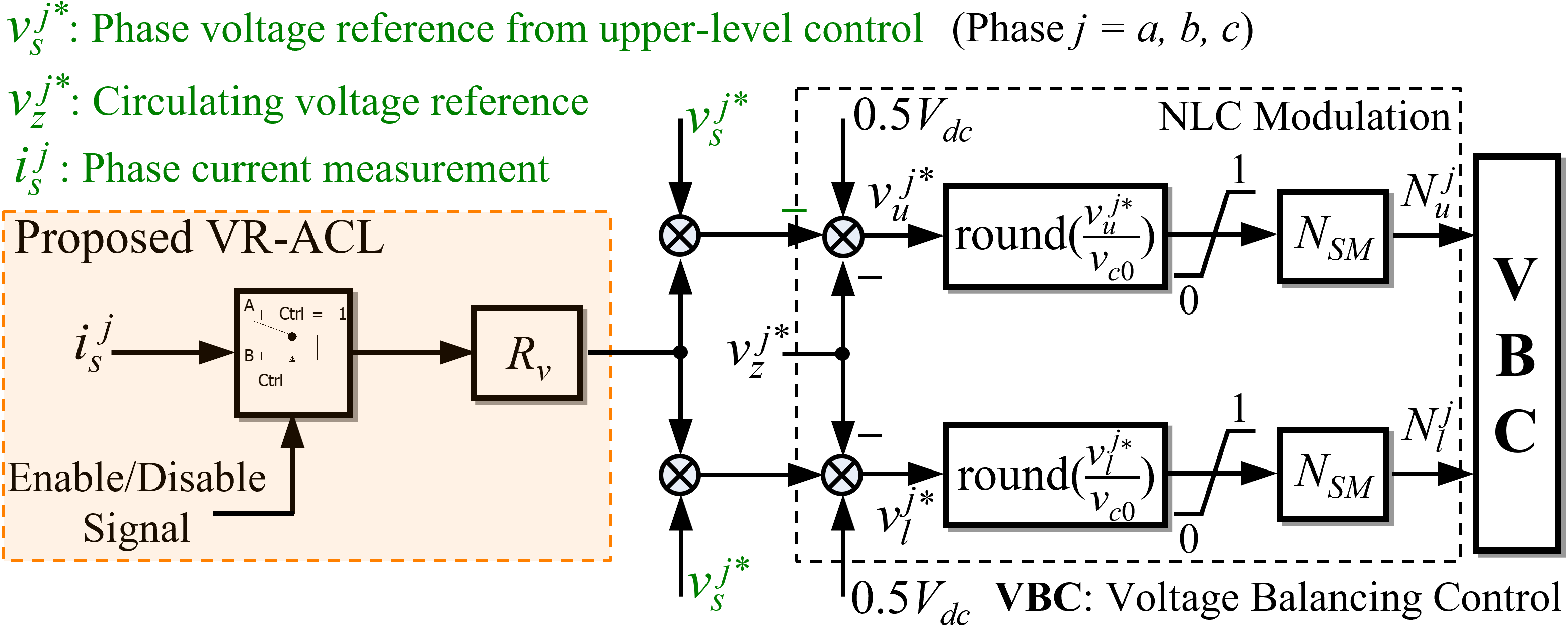}
    \caption{Control structure of VR-ACL control.}
    \label{fig: VR-ACL}
\end{figure}
Because the VAI-FCL control will not bypass all SMs, the non-bypassed SMs can be responsible for generating additional AC voltage to compensate for the voltage drop on the AC side. To achieve this, this work proposes a virtual resistance-based arm current-limiting (VR-ACL) control, which is depicted in Fig.~\ref{fig: VR-ACL}. Upon receiving the enabling signal, the VR-ACL control feed-forwards the phase current, $i_j$ ($j = a, b, c$), through a virtual resistor, $R_v$, to the AC voltage reference, $v_s^{j*}$. As a result, the voltage difference between the MMC and the AC grid is reduced during the periods when a certain amount of SMs are bypassed due to the activation of the VAI-FCL control. Because the VR-ACL control essentially modifies the modulation index of the AC voltage, its activation inevitably impacts the steady-state operation of the MMC; therefore, the VR-ACL control is only activated when the arm current exceeds 1.2 to 1.5 times its peak value, and it is deactivated immediately after the DCCB opening is triggered.

The design of the VR-ACL and its impact on the performance of the AFCL strategy represent a separate research effort, which will be discussed in future work.

\subsection{Design of TB-FCL control}

Because the TB-FCL control remains inactive during the steady-state operation of the MMC, its impact on the stability of the MTDC system is negligible; however, the design of the TB-FCL control must account for the transient behavior of the MMC during a DC fault to guarantee the efficacy of the TB-FCL  control.
\subsubsection{\texorpdfstring{$K_d$}{Kd}} 

As shown in Fig.~\ref{fig: TB-FCL}, $K_d$ determines the number of SMs to be bypassed. Unlike the VAI-FCL control, the TB-FCL control bypasses a fixed amount of SMs. For the TB-FCL control to function effectively, the amount of bypassed SMs must be greater than that bypassed by the VAI-FCL control. As an example, if the VAI-FCL has already bypassed $80\%$ of the SMs following a fault, and the TB-FCL is subsequently triggered with $K_d$ set to 0.7, then the SM discharge could become even more severe after the TB-FCL control is activated. 

In this paper, $K_d$ is set to one, thereby yielding the DC current behavior as characterized by \eqref{eq: Idc Solution 3}. This facilitates a complete bypass of all SMs and the fastest decay of the DC fault current. In practical applications, the minimum effective value of $K_d$ can be determined based on offline simulations of the MMC's discharge under the most severe DC fault. It should be noted that, as the TB-FCL control requires bypassing more SMs than the VAI-FCL to be effective in limiting the fault current peak, it results in a further increase in the arm current. Consequently, the optimal selection of the parameter $K_d$ must be coordinated with the design of the VR-ACL or other arm current-limiting methods.

\subsubsection{\texorpdfstring{$I_{max}$}{Imax}} 

$I_{max}$ determines when the TB-FCL control is activated. As discussed above, TB-FCL control can lead to a more severe overcurrent than the VAI-FCL under strong grid conditions. As a result, it is more suitable for TB-FCL control to perform as a backup protection strategy, and the activation of TB-FCL control can be appropriately delayed to the latest possible moment. Considering the control delay effect on triggering the bypass operation of the SMs, the value of $I_{max}$ is typically set at approximately $90\%-95\% $ of the DC current-blocking threshold. In addition, $I_{max}$ can also be determined based on the short-circuit current-breaking rating of a DCCB, $I_{brk}$. As an example, for a DCCB located on one end of a faulted line, if it is located along the discharge paths of $n$ MMCs, $I_{max}$ can be selected as $I_{brk}/n$.

\subsubsection{\texorpdfstring{$I_{min}$}{Imin}} 
$I_{min}$, together with $I_{max}$, determines the duration of the TB-FCL control. As discussed, the TB-FCL control leads to a more severe arm overcurrent than that induced by the VAI-FCL under strong grid conditions. In addition, when all the SMs are bypassed ($ K_d = 1 $ and $ D_\Delta = 0 $), the MMC loses its ability to emulate virtual resistance on the AC side through the VR-ACL, indicating no capability to limit arm overcurrents. As a consequence, the TB-FCL should be implemented for the shortest possible duration to minimize its negative effects, and $I_{min}$ can be selected as a few percentages less than $I_{max}$. 

\section{Performance Validation}
\label{Sec: V}
\subsection{Simulation Setup}

To validate the effectiveness of the proposed AFCL strategy for non-blocking DC fault ride-through of MMCs, this section presents two case studies in which two typical MTDC system layouts are considered: a radial topology and a meshed topology. 
The MTDC system is in a bipolar configuration, and all HVDC converters are modeled as HB-MMCs. The development of EMT models for the MMCs follows the guidelines outlined in \cite{CIGRE_B4_57}, whereas the frequency-dependent cable and the overhead transmission line model are modeled following the parameters and configuration given in \cite{AAPowerLink} and \cite{Li2019}. To more clearly demonstrate the effectiveness of the FCL, the case studies are settled as follows: 1) The SM blocking strategy follows the guidelines outlined in \cite{Willem-DCCB-Sizing} and \cite{CIGRE_B4_57}---that is, the SM blocking is activated when the DC output current of the MMC exceeds the DC current-blocking threshold, which is 2 pu of the rated DC current; 2) Hybrid DCCBs are employed in all scenarios to interrupt fault currents, with their activation configured to 6 milliseconds after fault inception; 3) A $200~\mu$s control delay is included in the MMC in each case. This helps to see the limitation of the proposed AFCL strategy when its design is limited by control delay-related stability constraints. Note that this is not considered in the existing literature.

 \begin{figure}[!b]
    \centering
    \includegraphics[width=0.5\textwidth]{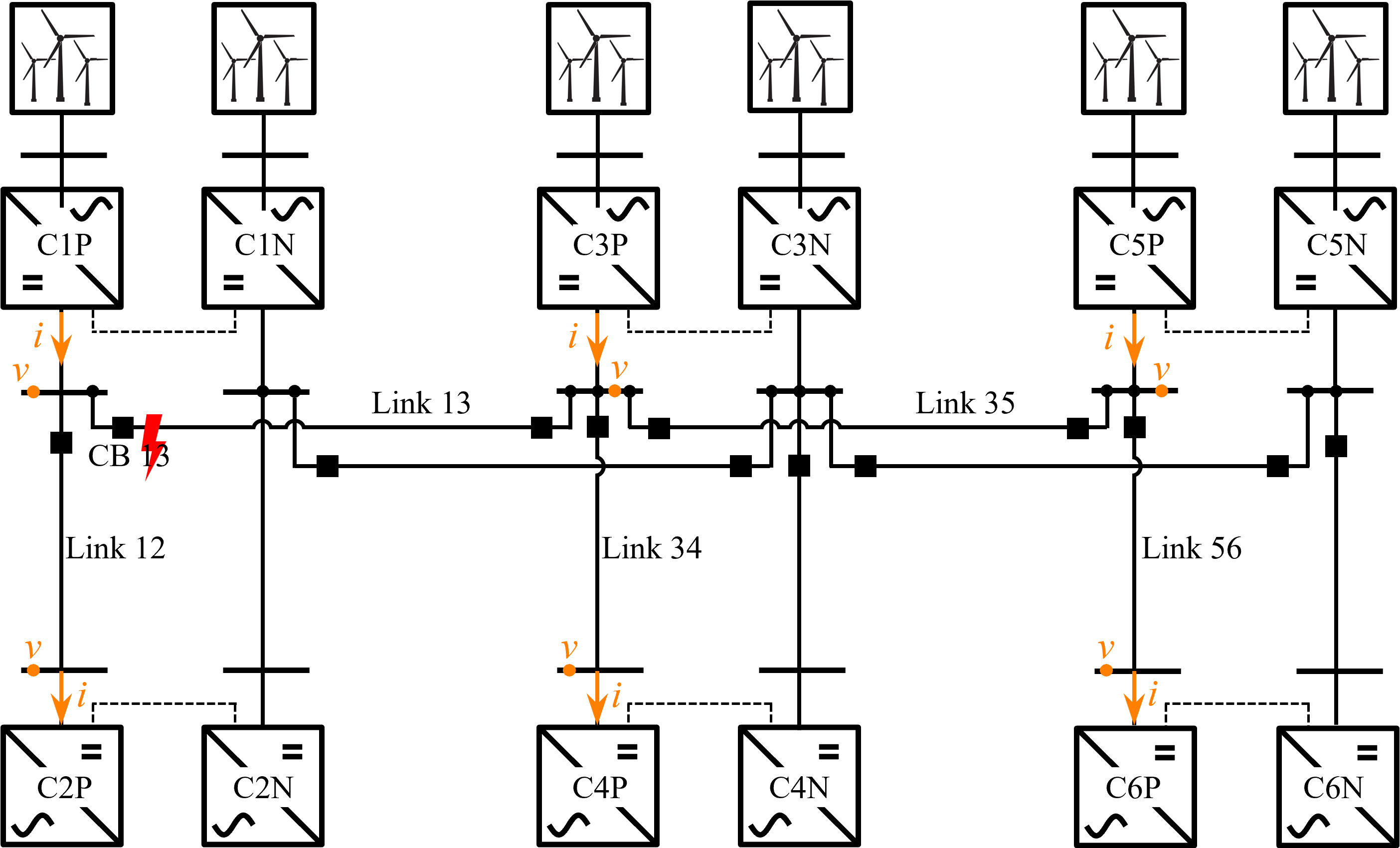}
    \caption{Six-terminal radial HVDC system. The measurement points are indicated in orange.}
    \label{fig: System Configuration Case 1}
\end{figure}

\subsection{Case 1: Radial MTDC-Based Wind Integration}

\subsubsection{Simulation Description} In the MTDC-based wind integration, the most commonly employed topology is the radial configuration, where HVDC substations at the sending terminals are interconnected through short transmission lines (e.g., 75 km). 
Due to the short distance of the interlinking cable, if a fault occurs near the sending terminals, the propagation of a DC fault transient can result in the blocking of multiple MMC stations at the sending terminals.

The six-terminal radial system shown in Fig. \ref{fig: System Configuration Case 1} is used for validation purposes. C-1, C-3, and C-5 are assumed to be stations at the sending end that integrate three wind power plants, and they supply power to stations C-2, C-4, and C-6 through export cables. The radial system consists of two links of 75 km (links 13 and 35), one link of 225 km (link 56), and two links of 150 km (links 12 and 34). A permanent pole-to-ground fault is applied to the positive pole of the system in link 13 at zero distance from CB13 at $t = 4$ s. 
The system parameters are tabulated in Table~\ref{Tab: MMC Parameter in Case 1}, while the parameters of the proposed VAI-FCL and the TB-FCL are tabulated in Table~\ref{Tab: AFCL}. 

\begin{table}[!t]
\centering
\caption{Parameters of Six-Terminal Radial HVDC}
\scalebox{0.8}{
\begin{tabular}{cccccccc} 
\toprule
\begin{tabular}[c]{@{}c@{}}\textbf{Parameter}\end{tabular} &
\begin{tabular}[c]{@{}c@{}}\textbf{C-1}\end{tabular} & 
\begin{tabular}[c]{@{}c@{}}\textbf{C-2}\end{tabular} & 
\begin{tabular}[c]{@{}c@{}}\textbf{C-3}\end{tabular} &
\begin{tabular}[c]{@{}c@{}}\textbf{C-4}\end{tabular} &
\begin{tabular}[c]{@{}c@{}}\textbf{C-5}\end{tabular} &
\begin{tabular}[c]{@{}c@{}}\textbf{C-6}\end{tabular} &\\
\midrule
DC-side voltage (kV)         &  $\pm$ 525                   &   $\pm$ 525        &   $\pm$ 525             &   $\pm$ 525   &   $\pm$ 525              &   $\pm$ 525 
\\
Rated capacity (MVA)     & 2000        &  2000                    &  2000         &  2000     & 2000        &  2000 
\\
Tx valve-side voltage (kV)    & 275     &  275                &   275     &  275    &  275       &  275
\\   
Tx grid-side voltage (kV)    & 66     &  400                &  66    &  400  &  66      &  400
\\      
Arm reactor (mH)      & 50          &  50                 &  50          &  50  &  50                 &  50 
\\                                
SM capacitance (mF)    & 15          &  15         &  15     &  15 &  15        &  15
\\
Control mode                  &  V/F                 & $ V_{dc}/Q $                 &  V/F                 &  $V_{dc} /Q$   &  V/F                 &  $V_{dc} /Q$   
\\
DC cur. blk. thresh. (kA)               &  3.81          & 3.81             & 3.81                  &  3.81     & 3.81                 & 3.81     
\\
Series reactor w/ DCCB (mH)               &  50          &  50             &  50                &   50     &  50               &  50  
\\
\bottomrule
\end{tabular}
}
\label{Tab: MMC Parameter in Case 1}
\end{table}

\begin{table}[!t]
\centering
\caption{Design of AFCL Control Strategy}
\scalebox{0.9}{
\begin{tabular}{cccc} 
\toprule
\begin{tabular}[c]{@{}c@{}}\textbf{Parameter}\end{tabular} &
\begin{tabular}[c]{@{}c@{}}\textbf{Symbol}\end{tabular} & 
\begin{tabular}[c]{@{}c@{}}\textbf{Value}\end{tabular} & 
\\
\midrule
Cutoff frequency of LPF            & $f_{c}$               & 5 Hz       
\\
Virtual arm reactor in with-delay case        & $L_v$               & 0.1 pu
\\     
Hysteresis on \& off point  for each MMC                   & $I_{max}$ \&  $I_{min}$             &  3.6 \& 3.5 kA
\\
Percentage of bypassed SMs  & $K_d$               & 100\%
\\
Virtual AC-side resistance  & $R_v$               & 3 pu
\\
\bottomrule
\end{tabular}
}
\label{Tab: AFCL}
\end{table}
\subsubsection{Simulated Responses}
\begin{figure}[!t]
    \centering
    \begin{subfigure}[b]{0.995\linewidth}
        \centering
        \captionsetup{skip= 1 pt} 
        \includegraphics[width=\linewidth]{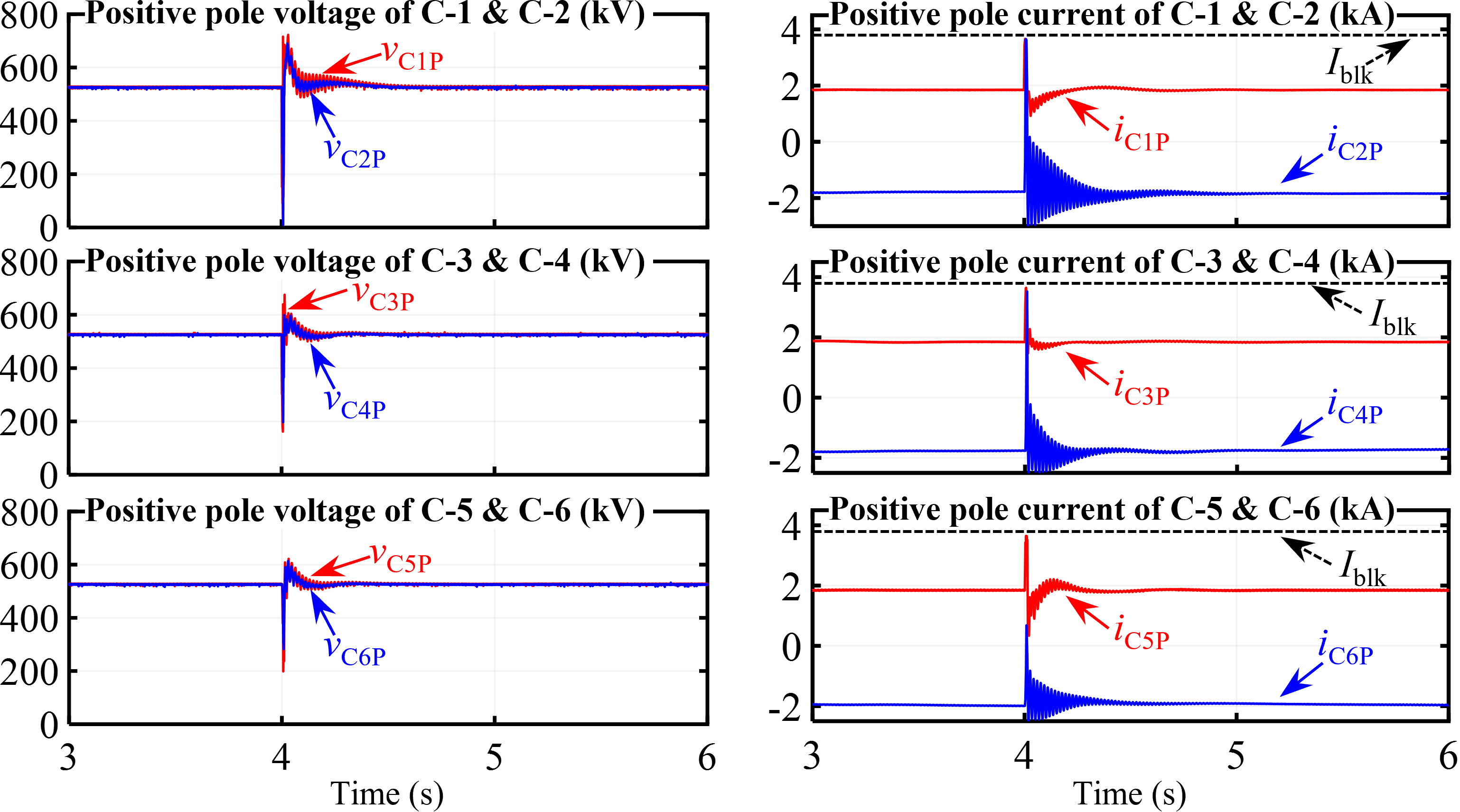}
        \caption{}
        \label{fig: Case 1 Voltage and Current}
        \vspace{0.5em} 
    \end{subfigure}
    \begin{subfigure}[b]{0.995\linewidth}
        \centering
        \captionsetup{skip= 1 pt} 
        \includegraphics[width=\linewidth]{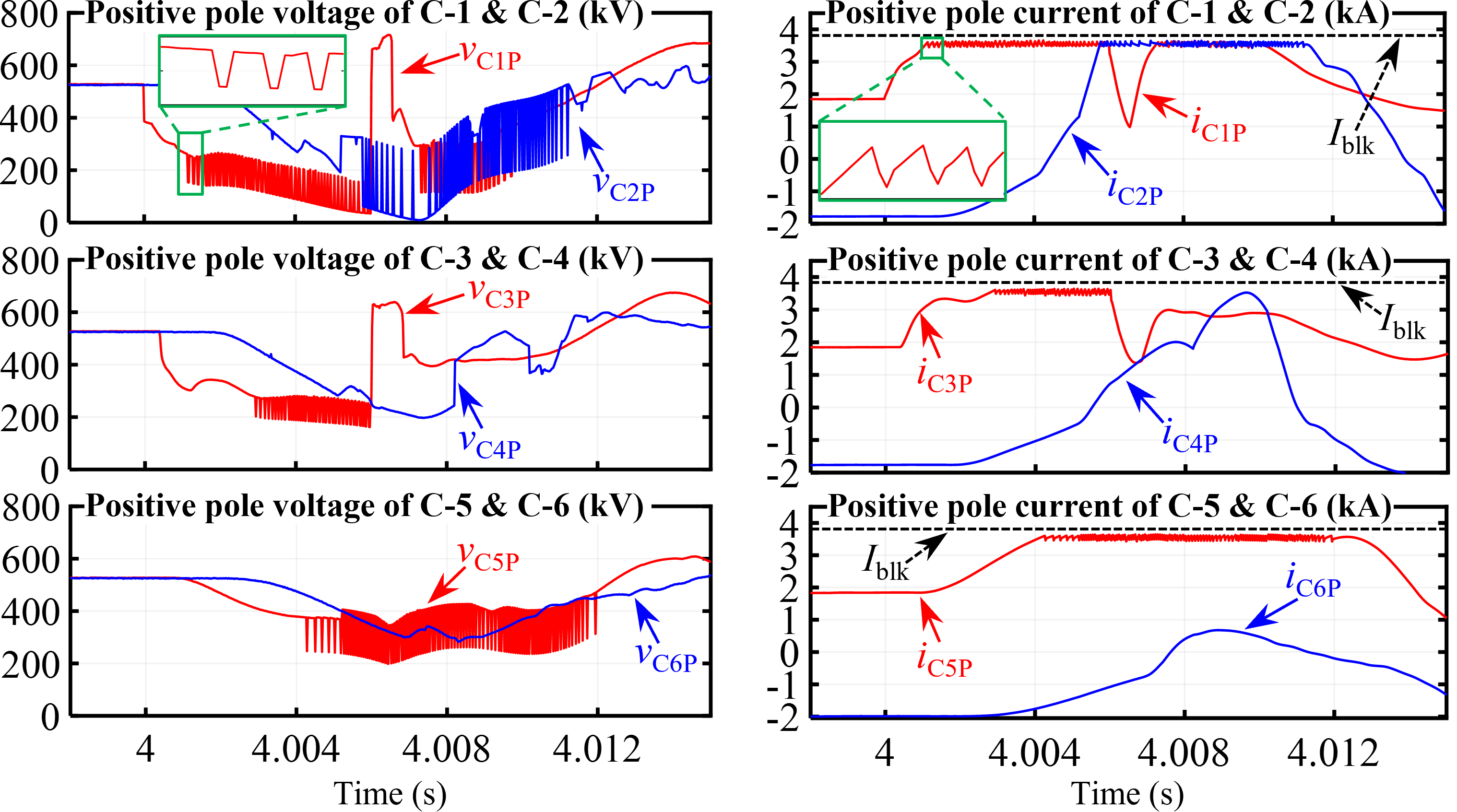}
        \caption{}
        \label{fig: Case 1 Zoom-in Voltage and Current}
    \end{subfigure}
    \vspace{-1.5em}
    \caption{Simulated responses of: (a) positive pole voltages and currents of MMCs C-1 to C-6; (b) zoom-in view of (a).}
    \label{fig: Simulation Results of Case 1}
\end{figure}

The simulated DC voltage and DC current responses of the six terminals are presented in Fig. \ref{fig: Simulation Results of Case 1}, where Fig. \ref{fig: Simulation Results of Case 1}(b) provides a zoomed-in view of Fig. \ref{fig: Simulation Results of Case 1}(a) during the first 15 ms after the fault occurs. 
 As shown in Fig. \ref{fig: Simulation Results of Case 1}(a), the DC voltage and current of each MMC recover to their steady-state values within approximately $150$ to $200$ ms. It is also evident that the DC current peak of each MMC does not exceed the DC current-blocking threshold of 3.81 kA (black dashed line in Fig. \ref{fig: Simulation Results of Case 1}(a)), indicating that no MMC is blocked during the DC fault. Also note that due to the time delay introduced by the pulse-width modulation and control, the actual peak value of the DC fault current slightly exceeds 3.6 kA. This indicates that when designing $I_{max}$, a sufficient current margin must be considered between $I_{max}$ and the DC current-blocking threshold.

To further analyze the DC current behavior of the MMCs, $i_{\mathrm{C}1\mathrm{P}}$ is considered as an example. It is observed that $i_{\mathrm{C}1\mathrm{P}}$ takes 1.14 ms to rise to 3.6 kA, during which the VAI-FCL functions. Note that without the VAI-FCL, the DC current would reach the blocking threshold of 3.8 kA within tens of microseconds, resulting in the tripping of C-1. Once the DC current reaches 3.6 kA, the upper threshold of the hysteresis band of the TB-FCL, the TB-FCL is activated and sets $D_\Delta = 0 $. As a result, all SMs in the C-1 are bypassed, leading to an immediate reduction in the DC fault current. When the DC current decreases to 3.5 kA, reaching the off-point of the hysteresis band, the TB-FCL is deactivated, and all SMs participate in switching. At this moment, because the DC fault on link 13 still exists, the SMs of C-1 continue to contribute fault current to the fault location, causing $i_{\mathrm{C}1\mathrm{P}}$ to reach 3.6 kA again at $t = 4.00128$ s, thereby retriggering the TB-FCL. This process repeats until the fault is cleared. 

One might notice that the DC current of C-1 immediately decays following the opening of the DCCB, after which it increases again, leading to the reactivation of the TB-FCL. The current increase is due to an unintended $LC$ transient oscillation triggered by the sudden change in the circuit topology upon opening the DCCB. Note that C-3 is also connected to the faulted line, so it shall experience transient oscillations following the opening of the DCCB as well. But the peak current remains below 3.6 kA during the oscillation, preventing the reactivation of the TB-FCL for C-3. The transient oscillations due to the $LC$ resonance in the DC grid have been extensively analyzed in \cite{Willem-Post-Fault-Recovery}, and the simulated responses demonstrate that the proposed AFCL strategy not only limits the DC current during a DC fault but also effectively suppresses the DC current after the fault is cleared by the DCCB.

\subsubsection{Arm Currents of MMC Without and With VR-ACL}
 \begin{figure}[!t]
    \centering
    \includegraphics[width=0.5\textwidth]{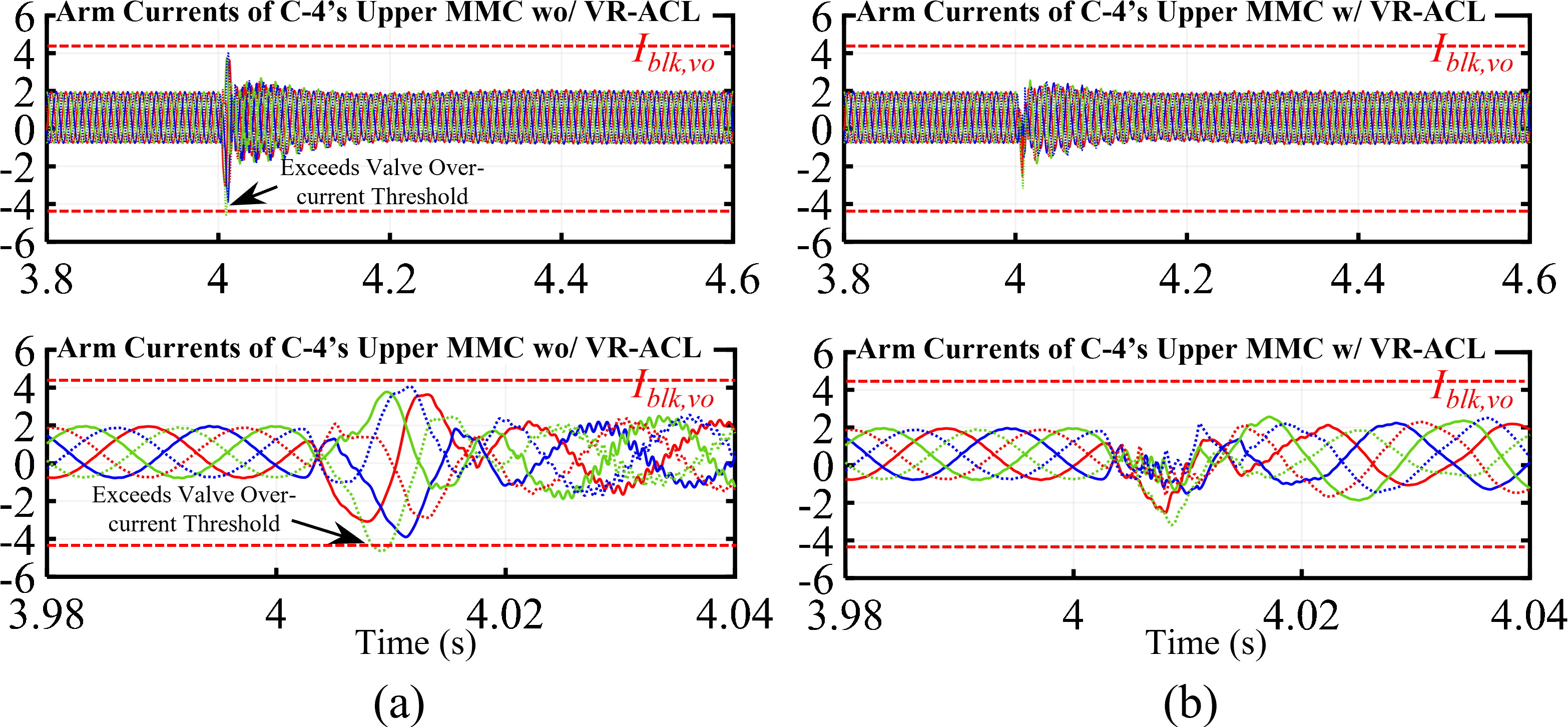}
    \caption{Simulated arm current responses of: MMC without VR-ACL; (b) MMC with VR-ACL.}
    \label{fig: Simulated Responses of Arm Current}
\end{figure}
As previously discussed, the AFCL strategy can induce arm overcurrents in MMCs that are connected to a strong grid. In the MTDC system illustrated in Fig.~\ref{fig: System Configuration Case 1}, C-1, C-3, and C-5—connected by wind plants—can be considered connected to weak grids; therefore, the arm currents of C-1, C-3, and C-5 typically do not experience overcurrents caused by the decrease of the output voltage of MMCs. In contrast, C-2, C-4, and C-6 are connected to land-based AC grids, the strength of which depends on the short-circuit ratio. In this case study, the grid behind the transformer of C-4 is purposely modeled as an ideal voltage source to emulate a strong grid condition. To validate the effectiveness of the VR-ACL during the application of the AFCL strategy, the simulated responses of the arm currents of C-4's upper MMC are compared with and without the involvement of the VR-ACL. 

Fig.~\ref{fig: Simulated Responses of Arm Current} illustrates the simulated responses of arm currents of the upper converter in C-4 during the fault, without and with the employment of the VR-ACL. The upper arm currents are represented by solid lines, and the lower arm currents are depicted by dashed lines. The red dashed line indicates the blocking threshold, $I_{blk, ov}$, of the valve overcurrent protection, which is 2 pu of the peak value of the arm current. In this study, the VR-ACL is configured to activate when the maximum value among the six arm currents exceeds 1.2 pu, resulting in its activation approximately 5 ms after the fault occurs for MMC C-4. It is evident that the peak arm current remains far below the valve overcurrent-blocking threshold (see Fig. \ref{fig: Simulated Responses of Arm Current}(b)), $I_{blk, ov}$, when using the proposed AFCL strategy, indicating that the valve overcurrent protection is not activated. If the VR-ACL is disabled, however, the MMC can experience larger arm overcurrent and consequently be blocked, even if the DC-side current is limited below the DC current-blocking threshold by the proposed AFCL strategy.

\subsection{Case 2: Four-Terminal Meshed HVDC Grid}
 \begin{figure}[!b]
    \centering
    \includegraphics[width=0.5\textwidth]{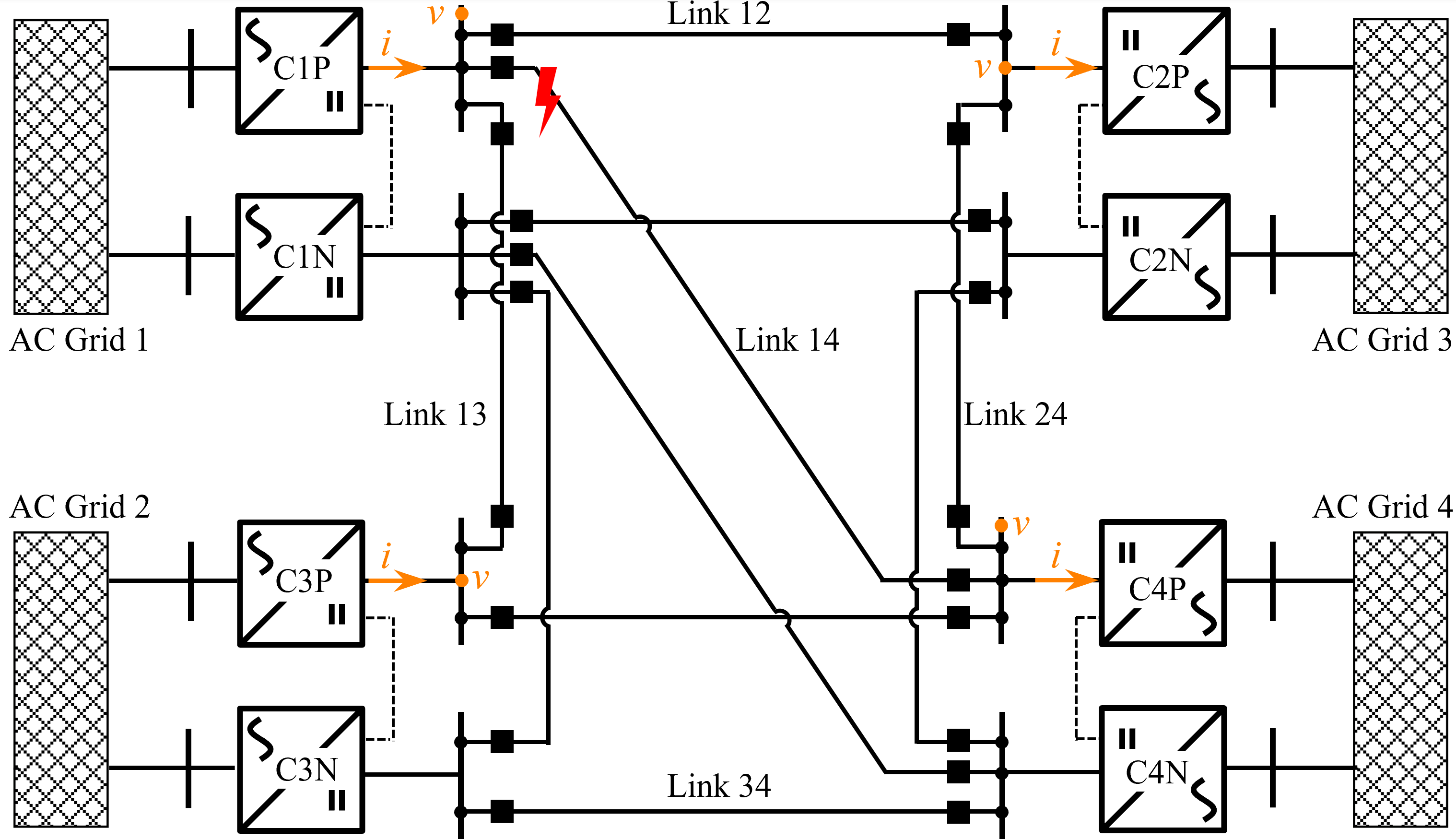}
    \caption{Four-terminal meshed HVDC grid. The measurement points are indicated in orange.}
    \label{fig: System Configuration Case 2}
\end{figure}
\subsubsection{Simulation Description} 
In a radial system (see Fig.~\ref{fig: System Configuration Case 1}), each MMC has a single path to the fault location; however, a meshed HVDC grid introduces multiple parallel paths, resulting in more complex fault current distributions and higher fault currents. This poses more challenges for the effectiveness of AFCL strategy. 

The four-terminal HVDC grid shown in Fig. \ref{fig: System Configuration Case 2} is used for validation purposes. The HVDC grid consists of two links of 200 km (links 13 and 14), one link of 150 km (link 24), and two links of 100 km (links 12 and 34). DCCBs are included at the end of each transmission line. The system parameters are tabulated in Table~\ref{Tab: MMC Parameter in Case 2}. It should be noted that the CLR connected in series with the DCCBs is rated at $50$ mH, which is significantly smaller than the values commonly reported in the existing literature on AFCL.
\begin{table}[!t]
\centering
\caption{Parameters of Four-Terminal Meshed HVDC}
\scalebox{0.9}{
\begin{tabular}{cccccc} 
\toprule
\begin{tabular}[c]{@{}c@{}}\textbf{Parameters}\end{tabular} &
\begin{tabular}[c]{@{}c@{}}\textbf{C-1}\end{tabular} & 
\begin{tabular}[c]{@{}c@{}}\textbf{C-2}\end{tabular} & 
\begin{tabular}[c]{@{}c@{}}\textbf{C-3}\end{tabular} &
\begin{tabular}[c]{@{}c@{}}\textbf{C-4}\end{tabular} &\\
\midrule
DC-side voltage (kV)         &  $\pm$ 525                  &   $\pm$ 525        &   $\pm$ 525             &   $\pm$ 525 
\\
Rated capacity (MVA)     & 2000        &  2000                   &  2000             &  2000      
\\ 
Tx valve-side voltage (kv)     & 275       &  275                  &  275        &  275  
\\ 
Tx grid-side voltage (kv)     & 400      &  400                  &  400        &  400
\\  
Arm reactor (mH)      & 50          &  50                 &  50          &  50 
\\                                
SM capacitance (mF)    & 15         &  15       &  15       &  15
\\
Control mode                  &  P-$V_{dc}$ droop                & P \& Q                &  P-$V_{dc}$ droop                     &  P \& Q    
\\
DC cur. blk. thresh. (kA)               &  3.81         & 3.81               & 3.81                             & 3.81      
\\
Series reactor w/ DCCB (mH)               &  50         &  50          &  50                &  50 
\\
\bottomrule
\end{tabular}
}
\label{Tab: MMC Parameter in Case 2}
\end{table}

A pole-to-ground DC fault in the positive pole is initiated in link 14 at zero distance from MMC C-1 at $t = 4$ s. The DCCB operating time and SM blocking strategy are designed the same as in Case 1. The parameters of the proposed VAI-FCL and the TB-FCL in this case are the same as those used in Case 1 (refer to Table~\ref{Tab: AFCL}). 
\begin{figure}[!t]
    \centering
    \begin{subfigure}[b]{0.995\linewidth}
        \centering
        \captionsetup{skip= 1 pt} 
        \includegraphics[width=\linewidth]{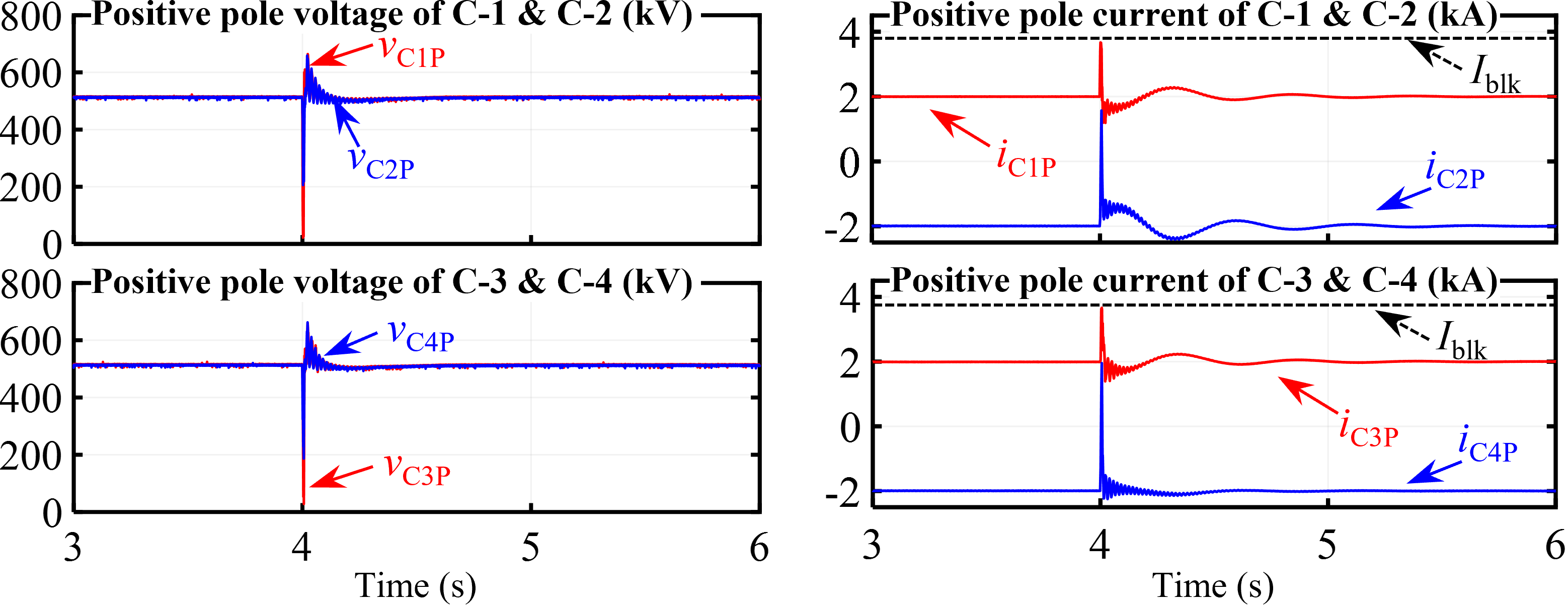}
        \caption{}
        \label{fig: Case 2 Voltage and Current}
        \vspace{0.5em} 
    \end{subfigure}
    \begin{subfigure}[b]{0.995\linewidth}
        \centering
        \captionsetup{skip= 1 pt} 
        \includegraphics[width=\linewidth]{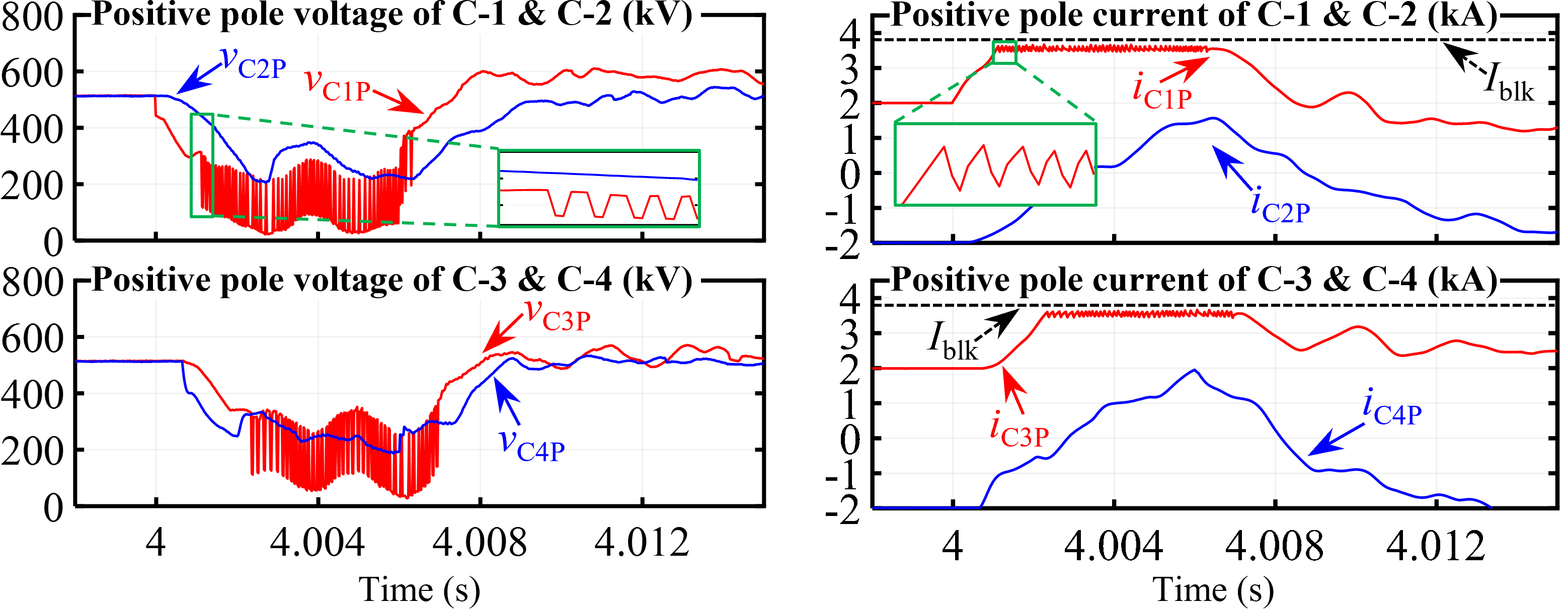}
        \caption{}
        \label{fig: Case 2 Zoom-in Voltage and Current}
    \end{subfigure}
    \vspace{-1.5em}
    \caption{Simulated responses of: (a) pole voltages and current of MMCs C-1 to C-4; (b) zoom-in view of (a).}
    \label{fig: Simulation Results of Case 2}
    \vspace{-1.5em} 
\end{figure}

\subsubsection{Simulated Responses} 
To save space, only the simulated DC voltage and current responses of each MMC are presented in this subsection, as shown in Fig. \ref{fig: Simulation Results of Case 2}. As observed, the DC current responses of all four MMCs remain below the DC current-blocking threshold during the DC faults, demonstrating the successful fault ride-through of the meshed HVDC grid. In particular, for C-1 and C-3, where the TB-FCL is activated, the DC current is effectively limited to approximately 3.6 kA. Once the DCCBs at the two ends of link 14 open to interrupt the fault current, the discharge paths of the four MMCs are eliminated, resulting in a gradual decrease in the DC current. Subsequently, the DC current returns to its steady-state value. For C-2 and C-4, because the DC current never reaches 3.6 kA before the fault clearance, the current limiting during the fault is solely performed by the VAI-FCL.

\section{Conclusion}
To enable non-blocking fault ride-through capability in HB-MMC-based MTDC systems during DC faults, a novel AFCL strategy is proposed in this paper, which includes VAI-FCL control and TB-FCL control. The VAI-FCL method introduces a virtual impedance into each arm of the MMC to effectively suppress the rate of rise of the fault current immediately following fault occurrence. In parallel, the TB-FCL control limits the peak value of the fault current by temporarily bypassing a predetermined number of SMs once the fault current exceeds a predefined threshold. A comprehensive design consideration and a systematic design approach for the proposed AFCL strategy are then presented. In addition, to mitigate the potential arm overcurrent caused by the AFCL strategy, VR-ACL is also introduced, which emulates a virtual AC-side resistor by utilizing inserted SMs. 

Two case studies have been carried out to demonstrate the effectiveness of the proposed AFCL approach. The simulation results validate the capability of the proposed AFCL strategy to enable non-blocking DC fault ride-through in HB-MMC across two representative MTDC system configurations. In addition, the CLRs placed in series with the DCCBs in both case studies are significantly smaller than those reported in existing literature.

\bibliographystyle{IEEEtran}
\bibliography{reference} 
\end{document}